\documentclass[10pt,journal,compsoc]{IEEEtran}
\usepackage{amssymb}
\usepackage{amsmath}
\makeatletter
\newif\if@restonecol

\usepackage[linesnumbered,ruled,vlined]{algorithm2e}
\usepackage{amsmath,amssymb,amsfonts,amsthm}
\usepackage{algpseudocode}
\usepackage{graphicx}
\usepackage{subfig}
\usepackage{booktabs}
\usepackage{threeparttable}

\newtheorem{theorem}{Theorem}
\newtheorem{lemma}{Lemma}
\usepackage{color}

 \usepackage{graphicx}
 \usepackage{graphicx}

\newcommand{\del}[1]{}
\begin{document}

\title{SecEL: Privacy-Preserving, Verifiable and Fault-Tolerant Edge Learning for Autonomous Vehicles}

\author{Jiasi~Weng,~
        Jian~Weng,~\IEEEmembership{Member,~IEEE,}
        Yue~Zhang,~
        Ming~Li~
        \IEEEcompsocitemizethanks{\IEEEcompsocthanksitem J. S. Weng, J. Weng, Y. Zhang and M. Luo are with the College of Information Science and Technology in Jinan University, and Guangdong/Guangzhou Key Laboratory of Data Security and Privacy Preserving, Guangzhou 510632, China. \protect\\Jian Weng is the corresponding author.}
        \protect\\
E-mail: cryptjweng@gmail.com
}

\IEEEtitleabstractindextext{%
\begin{abstract}
Mobile edge computing (MEC) is an emerging technology to transform the cloud-based computing services into the edge-based ones.
Autonomous vehicular network (AVNET), as one of the most promising applications of MEC, can feature edge learning and communication techniques, improving the safety for autonomous vehicles (AVs).
This paper focuses on the edge learning in AVNET, where AVs at the edge of the network share model parameters instead of data in a distributed manner, and an aggregator (e.g., a base station) aggregates parameters from AVs and at the end obtains a trained model.
Despite promising, security issues, such as data leakage, computing integrity invasion and fault connection in existing edge learning cases are not considered fully.
To the best of our knowledge, there lacks an effective scheme simultaneously covering the foregoing security issues.
Therefore, we propose \textit{SecEL}, a privacy-preserving, verifiable and fault-tolerant scheme for edge learning in AVNET.
First, we leverage the primitive of bivariate polynomial-based secret sharing to encrypt model parameters by one-time padding.
Second, we use homomorphic authenticator based on message authentication code to support verifiable computation.
Third, we mitigate the computation failure problem caused by fault connection.
%
%
Last, we simulate and evaluate SecEL in terms of time cost, throughput and classification accuracy.
The experiment results demonstrate the effectiveness of SecEL.
\end{abstract}
\begin{IEEEkeywords}
Edge learning, Autonomous vehicles, Privacy, Verification, Fault connection
\end{IEEEkeywords}}

\maketitle

\IEEEdisplaynontitleabstractindextext

\IEEEpeerreviewmaketitle

\section{Introduction}\label{sec:introduction}
\IEEEPARstart{I}{n} recent years, mobile edge computing (MEC) is regarded as a promising way to perform computation-consuming tasks by enabling both local processing and global coordination~\cite{mach2017mobile, mao2017survey, wang2018edge}.
Autonomous vehicular network (AVNET) becomes one of the popular applications of MEC, facilitating smart city development~\cite{kumar2016vehicular, zhang2017mobile}.
In particular, MEC nodes, such as smartphones or autonomous vehicles (AVs)' central controllers at the edge of the network, are allowed to provide sufficient computation resources and locally tackle computing tasks, and meanwhile, a server at an edge node (e.g., a base station) is responsible for global coordination~\cite{wang2018edge}.
These MEC nodes have powerful central processing units (CPUs) or graphics processing units (GPUs) and usually are close to vehicles, resulting in short response time and low transmission latency.
Therefore, more and more advanced technologies have been integrated into MEC.~\looseness=-1

Edge learning~\cite{wang2018edge, zhu2018towards} in AVNET is one of the advanced technologies of MEC, where AVs are able to collaboratively learn a high-accuracy machine learning model used to predict the road environment to improve vehicle safety. 
This collaboration paradigm enables an individual AV to obtain an accuracy-acceptable model, although the individual AV owns limited sensing data at local.
A practical case of edge learning in AVNET is the Tesla's machine learning model for auto-driving feeded with data from millions of Tesla vehicles~\cite{zhu2018towards}.
Following the popular network architecture for MEC~\cite{mcmahan2016communication}, in edge learning, there are multiple AVs being collaborative learners and a remote server being a parameter server.
Autonomous vehicles share model parameters which are trained on the local sensing data.
The server keeps a common machine learning model to train, by aggregating shared model parameters from multiple AVs and updating model parameters.
Individual AVs continually train on the same local sensing data using the updated parameters until the pre-defined loss threshold is reached.

\textbf{Problem Statement}:
Efforts have been taken to improve the performance of edge learning \cite{wang2018edge, zhu2018towards, samarakoon2018federated, kamp2018efficient}.
However, the privacy and security issues of edge learning may be not well-addressed:
(i) Sharing model parameters (instead of data) cannot protect data privacy due to the popular membership inference attacks~\cite{melis2018inference} and reconstruct attacks~\cite{hitaj2017deep} on the sharing model parameters.
These attacks can infer the specific vehicle-behaviour information implied in the local sensing data which natively causes privacy concern.
Recently, data privacy issue has attracted people's great attention, and international organization has announced data protection standard, such as GDPR \cite{marelli2018scrutinizing}.
Existing solutions employ traditional public key-based encryption to ensure privacy-preserving federated learning~\cite{aono2018privacy, weng2018deepchain}, but they do not consider the features of the MEC setting, e.g., resource-limited, bandwidth-hungry and dynamic AVNET.
Thus, a lightweight privacy-preserving scheme applied into edge learning is in need.
(ii) The aggregator at the remote server can be subject to various attacks on computing integrity~\cite{liu2017trojaning, bagdasaryan2018backdoor} leading to a manipulated model  misbehaving (refer it to Section \ref{sec:threats}).
It would disrupt the model accuracy and even poison AVs' sensing capability.
Existing privacy-preserving federated learning methods~\cite{aono2018privacy, bonawitz2017practical} protect data privacy but neglect data integrity.
Therefore, providing computing verification in a privacy-preserving scheme is inevitable for edge learning. In addition, considering the lightweight requirement in AVNET, the proposed scheme should enforce efficiency in the meantime.
(iii) Communication failure between MEC nodes and the remote server may happen and cause computation failure, since AVs usually have high mobility.
Concretely, some AV may lose its secret, making itself unable to obtain the returned learning result from the aggregator.
Here, computation failure can decrease the learning performance of this AV, which may cause the unexpected safety threats, which belongs to non-malicious failure mentioned in~\cite{kairouz2019advances}.
From this view, a proposed security scheme should particularly tolerate the fault connection problem.

In addition, we concretely focus on the Software-Defined Networking (SDN)-enabled network architecture for MEC, since it is recently standardized by 3GPP~\cite{laselva20183gpp} which promotes edge learning.
SDN is widely adopted and enables the capability of resources management.
An SDN-enabled server assembling SDN control modules can select the most suitable technology to ensure maximum-degree reliability of connection.
Most importantly, the SDN-enabled server can have a network-wide view of the connection state of communicating AVs~\cite{shantharama2018layback, peng2018sdn}, meaning that it can be aware of the event of fault connection, which is crucial for the proposed scheme.

\textbf{Our Contribution}:
To simultaneously overcome the aforementioned challenges, we propose a privacy-preserving, verifiable and fault-tolerant scheme for edge learning in AVNET, where AVs asynchronously share local model parameters to an SDN-enabled server in charge of aggregation.
The contributions are summarized as follows:
\begin{itemize}
\item We leverage the primitive of bivariate polynomial-based secret sharing and utilize the one-time padding method instead of asymmetric encryption to guarantee the secrecy of shared model parameters, which leads to a lightweight privacy-preserving scheme.
\item We seamlessly integrate the homomorphic authenticator technique into the proposed privacy-preserving scheme, which allows participating AVs to verify the computation results from the SDN-enabled server, avoiding getting manipulated results.
\item We enable SecEL to tolerate disconnected AVs, thereby improving the reliability of learning. Particularly, an honest but disconnected autonomous vehicle's secret can be recovered by a group of alive participants so that he can fluently obtain the computation results from the SDN-enabled server. Considering the rare bandwidth resources, there is no need to introduce the additional interaction to share secrets among the group of alive participants.
\item We lastly simulate SecEL in Python-based setting and evaluate its performance when it is applied into a popular model-learning case. The evaluation results demonstrate the effectiveness of SecEL in terms of time cost, throughput and learning accuracy.
\end{itemize}
\textbf{Roadmap}: The rest of our paper is organized as follows.
In Section~\ref{sec:preliminaries}, we briefly introduce the primitives involved in our paper, including federate learning, bivariate polynomial-based secret sharing and homomorphic authenticator. %
In addition, we also give a typical scenario of edge learning in AVNET.
In Section~\ref{sec:secRequire}, we present the security model of SecEL.
Subsequently,  we show the detailed design of SecEL over five phases in Section~\ref{sec:design}.
Afterwards, in Section~\ref{sec:analyze}, we give a security analysis of the proposed SecEL with respect to the security model.
We lastly demonstrate the performance results by evaluating SecEL in Section~\ref{sec:implementation}, survey the related work in Section~\ref{sec:related} and make a conclusion in Section~\ref{sec:conclusion}.
\section{Background}\label{sec:preliminaries}
In this section, we review the essential algorithm of federated learning. Afterwards, we introduce two applied cryptographic tools.
\subsection{Federated Learning}
Federated learning is a kind of distributed machine learning algorithm, where multiple machines and a central sever jointly solve the learning problem using distributed gradient descent techniques \cite{mcmahan2016communication}.
The learning problem is to obtain the optimized parameters by minimizing the loss function on training dataset.

Formally, let vector $\mathbf{D}$ and $\mathbf{w}$ represent training dataset and training parameters of the model, respectively.
Then, $L(\mathbf{w},\mathbf{D})$ defines the loss function on $\mathbf{D}$.
Therefore, when federated learning is adapted, $L_i(\mathbf{w}_i,\mathbf{D}_i)$ denotes the $i$-th machine's loss function, where $\mathbf{D}_i$ is maintained locally by the machine, and $L(\mathbf{w},\mathbf{D})$ presents the loss function at a central server.
Suppose $N$ machines exist and then the following formula will be hold:
$$L(\mathbf{w},\mathbf{D})=\frac{\sum_{i=1}^NL_i(\mathbf{w}_i,\mathbf{D}_i)}{N}.$$

To minimize $L(\mathbf{w},\mathbf{D})$, stochastic gradient descent method (SGD) is often used to find $$\mathbf{w}^* = \mathrm{arg\;min}\;L(\mathbf{w},\mathbf{D}).$$
Distributed SGD can be naturally derived from SGD.
It includes two major learning steps to achieve the learning problem, such as \textit{local update} and \textit{global aggregation}.
In local update, each machine preforms the machine learning task locally and trains a small-scale model.
In global aggregation, a central server collects the small-scale models together and updates parameters.

Concretely, it works as follows : (i) Each machine $i$ initializes the local model parameters $\mathbf{w}_i^{it}$ with the same randomized value, where $it=0$.
(ii) The step of local update is executed to update $\mathbf{w}_i^{it}$ by iteration via the following rule,
$$\mathbf{w}_i^{it+1}=\mathbf{w}_i^{it}-\eta \nabla L_i(\mathbf{w}_i^{it},\mathbf{D}_i),$$
where $\eta$ is the learning rate.
(iii) After one or multiple local updates, the step of global aggregation is performed by aggregating all machines' parameters (a.k.a. intermediate gradients) uploaded by the participating machines.
Herein, we let the frequency of local update be $15$, as it is recommended to be in range $[10,20]$ \cite{su2015experiments}.
That is, $$\mathbf{w}^{it+16}=\frac{1}{N}\sum_{i=1}^N\mathbf{w}_i^{it+15}.$$
After finishing the global aggregation step, $\mathbf{w}^{it+16}$ is broadcasted to the joining machines and they continue the local update step.
(iv) This learning process is repeated until the loss function at the cental server is minimized to
the predefined threshold.
Note that $\mathbf{w}_i^{it+15}$ should be encrypted by the individual participant before sharing, since they may expose the private information of the local sensing data by leveraging inference or reconstruct techniques~\cite{melis2018inference, hitaj2017deep}.

On the other hand, distributed SGD generally is classified into two categories: synchronous SGD and asynchronous SGD, depending on whether the process of uploading local parameters is synchronized.
Consider the features of AVNET environment including bandwidth-limitation and latency-sensitivity,
this paper employs weakly asynchronous SGD \cite{zhang2015staleness} which neutralizes the effect of synchronous and asynchronous SGD.
Particularly, it lets the central server wait for $s$ ($1\leq s \leq N$) number of participating machines who finish uploading their parameters, rather than all machines as synchronous SGD does.
Meanwhile, it also avoids the problem of stale gradients caused by asynchronous SGD.
\subsection{Bivariate Polynomial-based Secret Sharing}
Derived from Shamir's secret sharing scheme, bivariate polynomial-based secret sharing uses a bivariate polynomial to share secret instead of using a univariate one \cite{harn2017share}.
Suppose that a secret is hidden in the constant term of a bivariate polynomial. The secret owner can distribute two polynomials related the secret to a participant, rather than a point.
Thus, it distributes more information to participants than the univariate polynomial-based secret sharing does. Due to the additional information, a participant's lost secret is allowed to be rebuilt by other participants. In addition, each pair of AVs naturally share a common private key, which enables them to privately transmit messages.

Formally, a bivariate polynomial can be defined as $\textsf{F}(x,y)=\sum_{i=0}^{t-1}\sum_{j=0}^{t-1}a_{i,j}x^iy^j$ \textsf{mod} $p$, where $p$ is a prime larger than the secret which is hid in $\textsf{F}(0,0)$. Herein, the degree in both variate $x$ and $y$ is $t-1$. Our scheme uses a symmetric bivariate polynomial, in which coefficients $a_{i,j} = a_{j,i}$, which allows $\textsf{F}(x,y)$ to have a same threshold in variate $x$ and $y$ to recover the secret $\textsf{F}(0,0)$. In the contrary, in an ansymmetric bivariate polynomial, $a_{i,j}$ is not equal to $a_{j,i}$, so $\textsf{F}(x,y)$ does not exactly have threshold $t$.

\subsection{Homomorphic Authenticator}
Basically, homomorphic authenticator is used to indicate which input data is authenticated and how the input data should be correctly computed \cite{tran2016efficient, gennaro2013fully}.
A general homomorphic authenticator scheme includes a sequence of the following probabilistic polynomial time algorithms.

$\textsf{KeyGen}(1^\lambda)\rightarrow (\lambda)$. With a security parameter $\lambda$, it outputs authenticated key $s$ and public parameter $pp$.

$\textsf{Auth}(s, m, \tau)\rightarrow (\sigma)$. Taking the authenticated key $s$, a message $m$ and a label $\tau \in \{0,1\}^*$ as input, it generates an authenticated tag $\sigma$ for $m$ under $\tau$. 

$\textsf{Eval}$($pp$, $f$, $\boldsymbol{\sigma}$)$\rightarrow \sigma^*$. Given $pp$, a vector of authenticated tags $\boldsymbol{\sigma}=(\sigma_1,...,\sigma_n)$ and an arithmetic computation (it could be an arithmetic circuit), it outputs a new authenticated tag $\sigma^*$.
If $\sigma_i$ is the authenticated tag for $m_i$ ($i \in \{1,...,n\}$) as the output of some label program $P$ (to be introduced), then $\sigma^*$ authenticates $m^*$, where $m^*=f(m_1,...,m_n)$, as the output of the composed program $P^*=f(P_1,...,P_n)$.

$\textsf{Verf}(s, m^*, P, \sigma^*)\rightarrow \{\textsf{true}, \textsf{false}\}$. It is a deterministic algorithm to verify whether $\sigma^*$ authenticates $m^*$ by the labeled program $P$.

\noindent\textbf{Evaluation correctness} of the above scheme is indicated as follows.
If $P^*=f(P_1,..,P_n)$, $m^*=f(m_1,...,m_n)$, and $\sigma^* \leftarrow \textsf{Eval}(pp, P^*)$, then $\textsf{true}\leftarrow \textsf{Verf}(s, m^*, P^*, \sigma^*)$.

\noindent\textbf{Remark.} Label $\tau$ and labeled program $P$ are the essential components for homomorphic authenticator.
They define which data is authenticated and how data is evaluated, respectively.
Hence, SecEL uses the round index to label the shared parameters and the labeled program is the sum operation on the shared parameters from participants.
A labeled program is represented as $P=(f, \tau_1,...,\tau_n)$.
The composed program is generated by  $P^*=f(P_1,...,P_n)$, where the inputs of $P^*$ correspondingly are the different labeled inputs of $P_1,...,P_n$.
\section{Scenario Model and Security Model}\label{sec:secRequire}
In this section, we firstly introduce the edge learning scenario in AVNET. Then, we discuss the basic assumptions around the security threats in the network scenario. Further, we present the security goals with respect to the threats.
\begin{figure}
\centering
\includegraphics[width=\columnwidth]{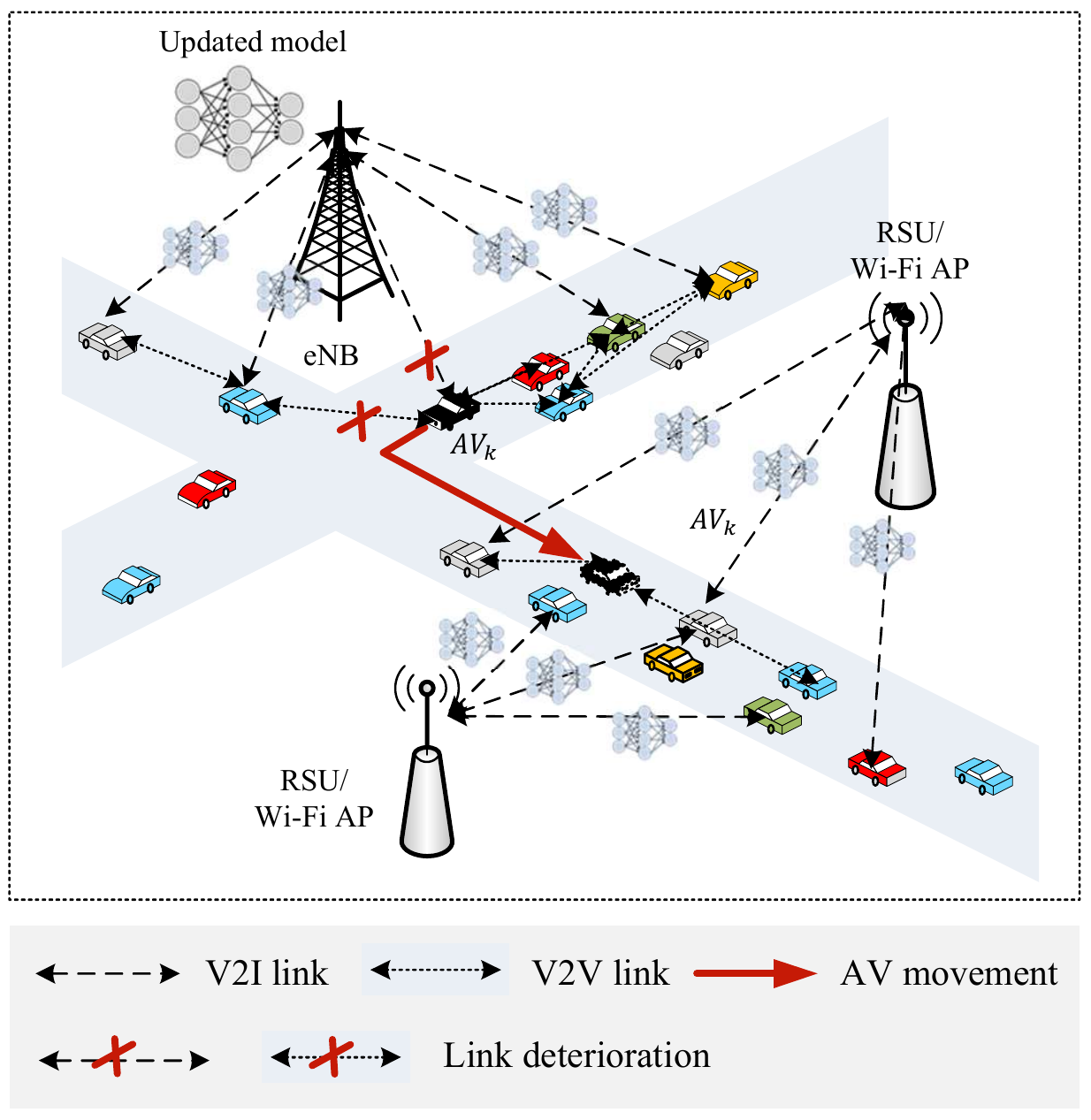}
\caption{Scenario model}
\label{scenario_model}
\end{figure}
\subsection{Scenario Model}
Fig.~\ref{scenario_model} shows an overview of the edge learning scenario in AVNET \cite{mcmahan2016communication}, where AVs collaborate to learn the vehicle environment for autonomous navigation with the help of edge nodes.
Through edge learning, AVs can \textit{indirectly} share sensing data to improve the prediction accuracy, i.e., the sharing model parameters which are trained on the local sensing data.
Specifically, edge nodes can be various base stations (BSs), including eNBs and RSUs/Wi-Fi APs.
AVs can connect with the closest BS via different communication technologies, such as dedicated short range communications (DSRC) and White-Fi.
V2I (vehicle to infrastructure) link presents the connection between AVs and BSs. V2V (vehicle to vehicle) link means the connection among vehicles.

Suppose that each AV has the small-scale sensing data due to the constrained sensing ability and fails to gain a high-accuracy prediction model.
Therefore, each AV can locally train a small-scale model on the local sensing data.
After training, AVs send intermediate gradients to an edge node (e.g., an eNB) that keeps and updates a common training model.
Then, AVs obtain the latest updated parameters from the edge node to update their local models.
The foregoing process repeats until the training loss is small enough.
For ease of presentation, it is defined that those AVs and the edge node are in a collaborative group, and they are participants.
Respectively, the entities outside the group are outsiders.\looseness=-1

Recall that we focus on the SDN-enabled MEC which is capable of resources management.
In this scenario, SDN control modules in edge nodes can help allocate network resources flexibly and efficiently for V2I link according to different requirements of communication qualities.
Note that an edge node's SDN control module maintains the global view of V2I and V2V links in its communication range.
Hence, the edge node can be aware of the link deterioration of AVs, if this AV moves out of the range of the connected AVs, like vehicle $\textsf{AV}_k$ in Fig.~\ref{scenario_model}.
Further, it is not difficult for BSs to identify which ciphertexts are from which AVs as shown in Section~\ref{sec:design}.
\subsection{Basic Assumptions\label{sec:threats}}
First, we assume that AVs are honest but curious, which means individual AVs can honestly participate in a collaborative learning but want to know more private information of other AVs like intermediate gradients.
That is because that there exists membership inference attacks~\cite{melis2018inference} and reconstruct attacks~\cite{hitaj2017deep}.
If AVs' intermediate gradients are exposed, they can be exploited to infer secret information of the corresponding local data.

Second, we assume that there exist fault AVs in the scenario. Fault AVs are those whose connections are invalid and they cannot response to other AVs inside a group. This is reasonable since AVs may offline due to various reasons.

Third, we assume that the edge node can be malicious, where he may manipulate or substitute the intermediate gradients collected from AVs and return incorrect updated parameters.
In this case, malicious attacks can be launched, such as trojaning attacks \cite{liu2017trojaning}.
Particularly, the adversarial edge node can inject trojaning parameters to the original model, through which the manipulated model would misbehave in a specific case.
To notice, prior work on the security of machine learning \cite{bonawitz2017practical} makes an assumption that the parameter server in their scheme (like the role of the edge node in this paper) is honest. On the contrary, we consider the worst case of our design, which is more realistic.

Last, we assume that there exist authenticated communication channels between entities (e.g. AVs, BSs).
Specifically, a symmetric authenticated encryption algorithm \cite{engels2011hummingbird} can be used to securely transmit data between any two parties who share a common secret key, in advance (to be explained in the Setup phase of Section~\ref{sec:design}).
\subsection{Security Goals}\label{sec:goals}
In this section, we claim our security goals of SecEL.
We will discuss the security issues, taking the features of AVNET into consideration, such as resource limitation, asynchronous communication and AVs' high mobility.

First of all, SecEL protects each AV's data privacy against other participants and outsiders. Specifically, individual AVs' model parameters cannot be leaked to anyone but itself. Only a threshold number of AVs can obtain the aggregated parameters.

Second, SecEL ensures the authenticity and correctness of computation at the aggregators or edge nodes.
It is noteworthy that in this resource-limited setting, the proposed verification method should be efficient.

Third, SecEL can tolerate the link-failed AVs and enforce the computation reliability.
Concretely, due to the high mobility of AVs and asynchronous communication, some AV can leave the communication range of other AVs and lose the secret shares (rather than the secret) which are given by other AVs in Setup phase (refer it to Section~\ref{sec:design}). This makes the AV fail to decrypt parameters returned by the aggregator, which may block his training process. Thus, SecEL guarantees that the AV's secret share can be recovered by other participants. For the sake of efficiency, this recovery process should not introduce the additional secret sharing round.
\begin{figure}[htbp]
\centering
\includegraphics[width=8cm, height=9cm]{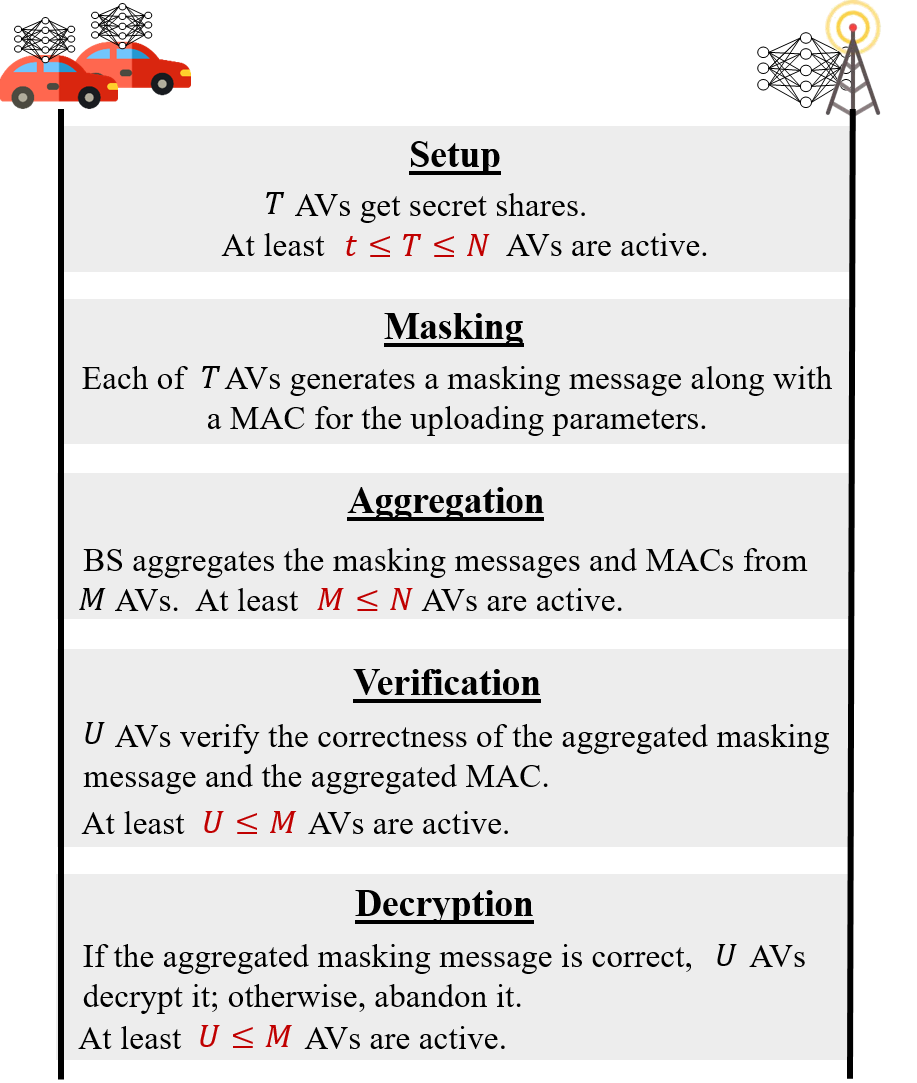}
\caption{Overview of SecEL. Herein, the number of active AVs in different phases (i.e., $[T],[M],[U]$) and their relationships are discussed in Fig.~\ref{fig:loss}.}
\label{overview}
\end{figure}
\section{SecEL Design}\label{sec:design}

\begin{table}[!t]
  \centering
  \caption{Summary of notations}\label{tab:notations}
  \begin{tabular}{|c|p{7cm}|}
        \hline
        \textbf{Notation} & \textbf{Description}\\ \hline
        $round$ & the version number of parameters on the BS\\ \hline
        $\textsf{AV}_i$ & the AV with identifier $i$\\ \hline
        $N$     & integer, the number of AVs participating in a round ($N$ varies in different rounds)\\ \hline
        $[N]$   & the $N$-length set that is \{$1,...,N$\}\\
                & (the similar meanings for $[M],[U],[T]$)\\ \hline
        $T$ & the size of $[T]$ (the similar meanings for $M,U$) \\ \hline
        $i\in [N]$ & the unique identifier of each AV \\ \hline
        $t$ & $f+1\leq t\leq N$, the threshold to reconstruct the secret \\ \hline
        $\textsf{F}_i(0,0)$ & also refer to as $V_i(0)$, a masking secret of $\textsf{AV}_i$\\\hline
        $s$ & an authenticated key \\ \hline
        $s_i$ & a partially authenticated key of AV $i$ and is a part of $s$ \\\hline
        $\mathcal{Z}_p$ & $\{1,2, ..., p-1\}$, $p$ is a secure prime larger than the secret \\ \hline
  \end{tabular}
\end{table}

As Fig. \ref{overview} shown, an overview of SecEL in one round which contains five phases, including Setup, Masking, Aggregation, Verification, Decryption.
A round here refers to a period when AVs share their intermediate gradients with the edge node and obtain aggregated parameters computed by the edge node.
Below we elaborate all the phases. To notice, we summarize the notations in Table \ref{tab:notations} for ease of description.
At last, we provide an alternative design to avoid running repeatedly the Setup phase in each round for sharing fresh secrets.
\subsection{Core Design}
This section presents the concrete design for five phases.

\textbf{Setup Phase.}
In this step, each AV selects a masking secret as well as a partially authenticated key, and shares them with other counterparts.
\begin{itemize}
  \item Masking secret. $\textsf{AV}_{i}$ hides the masking secret in $\textsf{F}_i(x,y)$ when $x=0, y=0$, where $\textsf{F}_i(x,y) \in \mathcal{Z}_p[x,y]$ with degree $t-1$ in variate $x$ and $y$ is a privately chosen symmetric bivariate polynomial.
  \item Authenticated key. $\textsf{AV}_{i}$ randomly selects key $s_i \in \mathcal{Z}_p$ which is called as partially authenticated key.
\end{itemize}

Afterwards, $\textsf{AV}_i$ distributes shares \{$\textsf{F}_i(x,j)$, $\textsf{F}_i(j,y)$\} of $\textsf{F}_i(0,0)$ and key $s_i$ to $\textsf{AV}_j$ ($j\neq i$ \& $j\in [N]$). For each AV, no less than $t\leq T\leq N$ other AVs are assumed receiving his secret shares.\looseness=-1

According to the properties of symmetric bivariate polynomial, we have the following insights: 

$\circ$ $\textsf{F}(0,0)=\sum_{i=1}^N\textsf{F}_i(0,0)$; 

$\circ$ $t$ number of $\textsf{F}_i(0,j)$ or $\textsf{F}_i(j,0)$ are sufficient to reconstruct $\textsf{F}_i(0,0)$; 

$\circ$ If any $\textsf{F}_i(0,j)$ owned by $\textsf{AV}_j$ is lost, it can be recovered by $t$ number of $\textsf{F}_i(q,j)$ ($q\in[T]$); the similar case for $\textsf{F}_i(j,0)$ holds.

$\circ$ $\textsf{AV}_{i}$ and $\textsf{AV}_j$ share a common secret key with respect to $\textsf{F}_i(x,y)$ (or $\textsf{F}_j(x,y)$), since $\textsf{F}_i(i,y)$ owned by $\textsf{AV}_{i}$ at $y=j$ equals to $\textsf{F}_i(j,y)$ owned by $\textsf{AV}_j$ at $y=i$.

\noindent\textbf{Refinement.} The above secret distribution can be bandwidth consuming (up to $\mathcal{O}(N^2t)$ communication complexity in total), which may be incompatible to the bandwidth-hungry setting.
To counter such a drawback, we deploy Algorithm \ref{algo:secretsharing} to distribute the secret with communication complexity $\mathcal{O}(Nt)$.
Algorithm \ref{algo:secretsharing} is borrowed from \cite{maramdynamic}, but we allow each AV to share his secret to $N-1$ counterparts rather than a portion of them in the first step.
Basically, Algorithm \ref{algo:secretsharing} contains two steps: (i) $N$ AVs individually choose a degree-$(t-1)$ univariate polynomial $V_i(x) \in \mathcal{Z}_p[x]$, where $i \in \{1,...,N\}$.
Then, each of them allocates shares of $V_i(0)$ to other $N-1$ AVs.
At the end of this step, each one $\textsf{AV}_i$ obtains the share $V(i)$ of $V(x)=\sum_{i=1}^{N}V_i(x)$. $V(i)$ is defined as $s_{v_i}$. (ii) Each $\textsf{AV}_i$ $(i\in [N])$ generates another degree-$(t-1)$ univariate polynomial $A_i(x) \in \mathcal{Z}_p[x]$ and $A_i(0)=s_{v_i}$. Then, $\textsf{AV}_i$ redistributes secret shares of $A_i(0)$ to all AVs, that is, $A_i(j)$ is distributed to $\textsf{AV}_j$ ($j\in [N]$). In this way, the foregoing $N$ number of $\{A_i(x)\}_{i\in [N]}$ with degree-$(t-1)$ encode the bivariate polynomial $\textsf{F}(x,y) \in \mathcal{Z}_p[x,y]$ with degree-$(t-1,t-1)$ and $\textsf{F}_i(x,i)=A_i(x)$. Besides, $\textsf{F}(0,0)=V(0)=\sum_{i=1}^{N}V_i(0)$.
$V(0)$ can be reconstructed after collecting at least $t$ shares $V(j)$ ($j\in [T]$). Similarly, $V(j)$ can be recovered by at least $t$ shares $A_j(q)$ ($q\in [T]$) because of $V(j)=A_j(0)$.

\begin{algorithm}
\caption{Each AV gets secret shares with communication complexity $\mathcal{O}(Nt)$}
\label{algo:secretsharing}
\LinesNumbered
\KwIn{$N$ AVs with identifier $i \in [N]$}
\KwOut{Each $\textsf{AV}_i$ ($i\in[N]$) possesses $s_{v_i}$ and $\{A_j(i)\}$ ($j\in[N]$)}
\underline{Step 1.}\\
\For{each $\textsf{AV}_i$ ($i \in [N]$)}{
choose a $(t-1)$-degree $V_i(x) \in \mathcal{Z}_p[x]$\;
    \textcolor[rgb]{0.6,0.6,0.6}{// keep $V_i(0)$ secretly}\;
    \For{each $\textsf{AV}_j$ ($j \in [N]$)}{
        compute $V_i(j)$ (mod $\mathcal{Z}_p[x]$)\;
    \textcolor[rgb]{0.6,0.6,0.6}{// distribute $V_i(j)$ ($j\neq i$) to $\textsf{AV}_j$}\;
    \textcolor[rgb]{0.6,0.6,0.6}{// keep $V_i(i)$ locally}\;
    }
}
\For{$i \in [N]$}{
    compute $\sum_{j=1}^{N}V_j(i)$\;
    $s_{v_i}=\sum_{j=1}^{N}V_j(i)$\;
}
\underline{Step 2.}\\
\For{each $\textsf{AV}_i$ ($i\in [N]$)}{
    generate a $(t-1)$-degree $A_i(x)\in \mathcal{Z}_p[x]$\;
    \textcolor[rgb]{0.6,0.6,0.6}{//where $A_i(0)=s_{v_i}$}\;
    \For{each $\textsf{AV}_j$ ($j\in [N]$)}{
        compute $A_i(j)$ (mod $\mathcal{Z}_p[x]$)\;
    \textcolor[rgb]{0.6,0.6,0.6}{// allocate $A_i(j)$ ($j\neq i$) to $\textsf{AV}_j$}\;
    }
}
\end{algorithm}

\textbf{Masking Phase.} AVs mask intermediate gradients and generate the corresponding authenticated MAC with the masking secret and authenticated key.
$\textsf{AV}_i$ first builds $\textsf{PRG}(V_i(0),round)$. Here, $\textsf{PRG}$ is the pseudorandom generator agreed among AVs.
After that, he computes $c_{1,i}=(\textsf{PRG}(V_i(0),round)+w_i)$ mod $\mathcal{Z}_p$ to mask parameters, where $w_i$ is the element of $\mathbf{w}_i$.
For each element, $\textsf{PRG}$ uses the distinct public nonce to generate random number.

To compute the second component, $\textsf{AV}_i$ generates a common authenticated key $s$ composed of $s_i$ $(i\in [N])$, i.e., $s=(\sum_{i=1}^{N}s_i)$ mod $\mathcal{Z}_p$.
Besides, $\textsf{AV}_i$ prepares $k_i=V_i(i)$ since he possesses $V_i(x)$. Based on $s$ and $k_i=V_i(i)$, $\textsf{AV}_i$ can compute $c_{2,i}=(\frac{\textsf{PRG}(k_i,round)-c_{1,i}}{s})$ mod $\mathcal{Z}_p$.
Then, AVs send their masking results \{$i$, ($c_{1,i},c_{2,i}$)\} ($i\in [N]$) to BS.

\noindent\textbf{Remark.} It is worth noting that $\textsf{AV}_i$'s $k_i$ can be recovered by at least $t$ AVs by providing $V_i(l)$ ($l\neq i$ \& $l\in [N]$).

\textbf{Aggregation Phase.} BS aggregates together the received ciphertexts in this phase.
BS receives AVs' ciphertext and records their identifier.
If some AV's ciphertext fails to submit, BS will note his identifier in the returned result because they have no contribution in this round.
Suppose that BS successfully receives $M$ AVs' ciphertext, $M\leq N$. BS separately aggregates two components of ciphertext: $c_1=\sum_{i=1}^{M}c_{1,i}$ and $c_2=\sum_{i=1}^{M}c_{2,i}$.
Next, it returns the aggregated result enclosing the failed AVs' identifiers, that is, \{$[N]\setminus[M],(c_1,c_2)$\}.

\textbf{Verification Phase.} Each AV verifies the correctness of aggregated results returned by BS. Suppose that $U$ ($U\geq t$) number of AVs receive results for keeping liveness \cite{maramdynamic}.
They execute a distributed, unpredictable and unbiased randomness algorithm \cite{syta2017scalable} to select an AV leader to collect shares from other AVs, and then recover secrets.
In this way, they collaboratively recover $k=\sum_{i=1}^{M}k_i$ which is used to generate $c_2=\sum_{i=1}^{M}c_{2,i}$ in Masking phase.
Recall that at least $t$ AVs can rebuild one of $\{k_i\}_{j\in [M]}$ and then $k$.
$\textsf{PRG}(k,round)$ can be computed with $k$.
Next, each of them verifies the correctness of $c_1$ and $c_2$ by identifying whether $\textsf{PRG}(k,round)$ is equal to $c_2\cdot s+c_1$.
If yes, they go into the next phase; otherwise, they reject the result.

\noindent\textbf{Remark.} Note that active AVs in this phase can privately recover the secrets against outsiders and the edge node.
The fact is that each pair of AVs naturally share a common secure key after the Setup phase, i.e., $\textsf{F}_i(i,j)$ or $\textsf{F}_j(j,i)$, between $\textsf{AV}_i$ and $\textsf{AV}_j$.
With this common key, each of them can securely transmit data by employing a symmetric authenticated encryption algorithm.
In doing so, there is no need to additionally run a key-agreement protocol to share a common key as done in work~\cite{bonawitz2017practical}.

\textbf{Decryption Phase.} AVs collaboratively unmask the correct ciphertext. Specifically, AVs $\in U$ further jointly reconstruct $\textsf{F}(0,0)$ (also $V(0)$) which is used to mask the intermediate messages. Note that at least $t$ AVs can collaboratively obtain $\{V_i(0)\}_{i\in[M]}$ and then $V(0)$. $\textsf{PRG}(V(0),round)$ can be computed with $V(0)$. After that, the aggregated message $\sum_{i=1}^Mw_i$ can be unmasked by $c_2-\textsf{PRG}(V(0),round)$. Then, each of AVs use $\sum_{i=1}^Mw_i$ to update the local parameters and the next round begins.

\noindent\textbf{Remark.} In the case of link deterioration, we assume that $\textsf{AV}_q$ ($q$ $\notin [T]$ but $\in [U]$) loses her share. Her secret still can be recovered only if there are at least $t$ AVs $\in [T]$ in the group. That is because $t$ AVs are sufficient to rebuild $\textsf{AV}_q$'s share of $\textsf{F}(0,0)$. We emphatically explain it in the following text.

\begin{figure}[htbp]
\centering
\includegraphics[width=0.8\columnwidth]{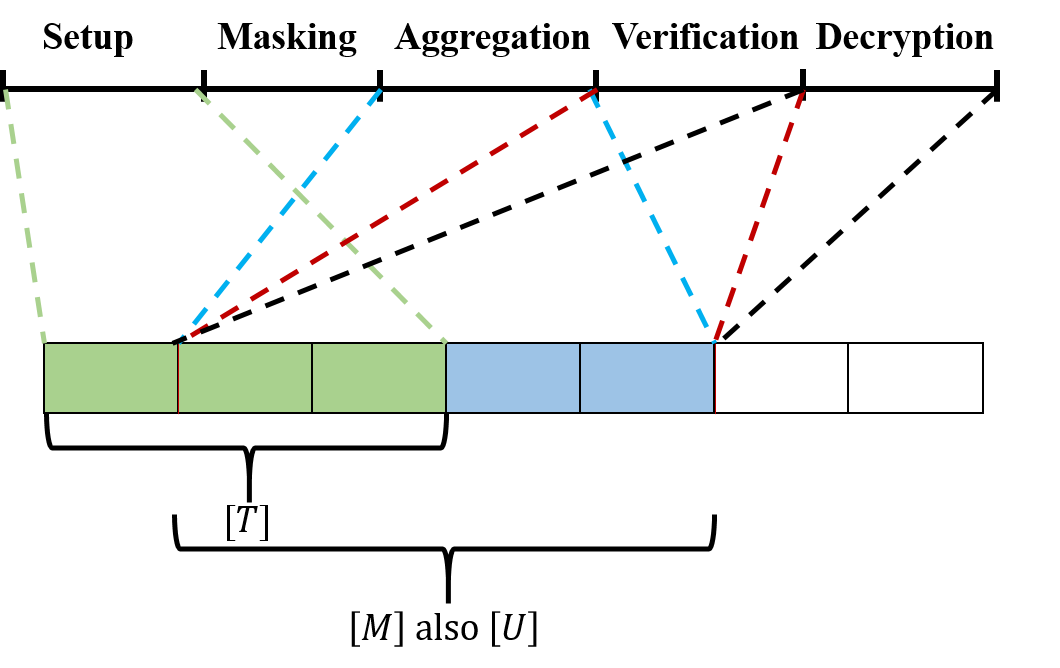}
\caption{Case of AVs' participation during a round. We use seven squares to represent seven AVs and suppose $t=3$. The dotted lines with different colors dedicate the participated AVs in respective phases, e.g., the green dotted lines appoint the three AVs (in green squares) participating in Setup phase, which means they have received shares. Note that AVs' identifiers belong to the corresponding sets mentioned in Fig.~\ref{overview}.}
\label{fig:loss}
\end{figure}

Once an honest AV loses secret, saying $V(q)$ (i.e. $\textsf{F}(0,q)$), no less than $t$ AVs who can provide shares $A_q(j)$ ($j\in [T]$) can help rebuild $V(q)$.
Here, we demonstrate the possible occurrence of the case of AVs' shares missing with the help of Fig.~\ref{fig:loss}.
Obviously, there are four AVs (due to four squares) contributing in Aggregation phase and they have to verify and decrypt the returned ciphertext in Verification phase and Decrypt phase, respectively.
However, two AVs (in the blue squares) have no shares, since they do not participate in Setup phase due to asynchronous communication.
Then, they need help from other three AVs (in the green squares) to reconstruct their shares.
In addition, the remain AVs (in the white squares) are not permitted to obtain the returned result since they do not make their contributions in Aggregation phase.
\subsection{Alternative Design}
For not rerunning the Setup phase, this section seeks to introduce an alternative design making shared secrets reusable in a secure manner.

The main idea is that each AV distributes encrypted shares to other AVs, where each share to some AV is encrypted under the public key of this AV, and meanwhile the respective secret is recovered without leaking any information of shares. Here, we denote a pair of private and public key for $\textsf{AV}_i$, i.e., ($sk_i, G^{sk_i}$), where $G$ is a public parameter and a generator in a multiplicatively cyclic group $\mathcal{G}$. We assume that they are secretly generated and public to all AVs. Then, the modified workflows for each step are shown as follows.\looseness=-1

\textbf{Setup phase.} $\textsf{AV}_i$ allocates share $G^{sk_jV_i(j)}$ and $G^{sk_jA_i(j)}$ to $\textsf{AV}_j$ (instead of $V_i(j)$ and $A_i(j)$ in Algorithm~\ref{algo:secretsharing}), which only occurs at the first round.
However, authenticated key $s_i$ still needs to be freshly chosen and sent at each round, which leads to comparatively negligible cost.
In this case, secret $G^{V_i(0)}$ can be later reconstructed with at least $t$ shares $G^{V_i(q)}$ provided by $\textsf{AV}_q$ who knows private key $sk_q$ ($q \in [T]$).
Similarly, secret $G^{A_i(0)}$ can be reconstructed by at least $t$ shares $G^{A_i(q)}$ ($q \in [T]$).

\textbf{Masking phase.} $\textsf{AV}_i$ generates masking results $\{i, (G^{c_{1,i}}, G^{c_{2,i}})\}$.
Note that this phase uses fresh key $s=\sum_{i=1}^{N}s_i$ (mod $\mathcal{Z}_p$) each round.

\textbf{Aggregation phase.} $c_1$ is computed by $\prod_{i=1}^{M}c_{1,i}$ and $c_2=\prod_{i=1}^{M}c_{2,i}$.

\textbf{Verification phase.} AVs verify whether $c_2^s \cdot c_1$ is equal to $G^{\sum_{i=1}^M\textsf{PRG}(k_i, round)}$, where $[M]$ AVs provide $G^{\textsf{PRG}(k_i, round)}$.

\textbf{Decryption phase.} $G^{\sum_{i=1}^Mw_i}$ can be computed by $c_2\cdot 1/G^{\textsf{PRG}(V(0),round)}$ due to $\{G^{\textsf{PRG}(V_i(0), round)}\}_{i\in[M]}$ can be recovered and then $G^{\textsf{PRG}(V(0),round)}$ can be calculated.
Here, $\textsf{PRG}$ needs to satisfy the following properties.
It satisfies $\textsf{PRG}(key_1+key_2,round)=\textsf{PRG}(key_1,round) + \textsf{PRG}(key_2,round)$ and $cst\cdot \textsf{PRG}(key,round)=\textsf{PRG}(cst\cdot key,round)$, where $cst$ is a constant~\cite{naor1999distributed}.
Finally, $\sum_{i=1}^Mw_i$ can be obtained by using the Baby-Step-Giant-Step
(BSGS) algorithm which can be realized with a reasonable cost for any integer no larger than $32$~bits~\cite{smart2003cryptography}.
\section{Security Analysis}\label{sec:analyze}
In this section, we give the security analysis with respect to the security goals mentioned in Section \ref{sec:goals}. Recall that SecEL devotes to protect data privacy, guarantee computation integrity and ensure reliability. However, whether these issues can be addressed depends on if the primitive of the symmetric bivariate polynomial is secure enough.
Therefore, we first give the analysis that symmetric bivariate polynomial we used is secure (Theorem~\ref{thm:1}), and then we present the security analysis for the foregoing security goals.

\begin{theorem}\label{thm:1}
Symmetric bivariate polynomial $\textsf{F}(x,y)$ is a $(t,N)$ secure secret sharing scheme.
\end{theorem}

\noindent\textbf{Proof.} It needs to be satisfied that only $t\leq N$ participants can reconstruct the secret, thus we prove less than $t$ number of participants are unable to gain the secret.

\noindent Note that the selected $\textsf{F}(x,y)$ has total $t+\frac{t(t-1)}{2}$ coefficients. Suppose that there exist $t-1$ participants colluding. $i$ has two ($t-1$)-degree univariate polynomials \{$\textsf{F}(x,i)$, $\textsf{F}(i,y)$\} (note: $\textsf{F}(j,i)=F(i,j)$) and can build $t$ linear independent equations. Then, $t-1$ colluded participants totally gain $t(t-1)$ linear independent equations. At the same time, those $t-1$ colluded participants, with each other, share $\frac{(t-1)(t-2)}{2}$ points of the bivariate polynomial $\textsf{F}(x,y)$. Thus, they finally can build $t(t-1)-\frac{(t-1)(t-2)}{2}$ linear independent equations at total to solve the bivariate polynomial $\textsf{F}(x,y)$. Since $t+\frac{t(t-1)}{2}$ is larger than $t(t-1)-\frac{(t-1)(t-2)}{2}$, $t-1$ colluded participants cannot reconstruct $\textsf{F}(x,y)$ and gain the secret.

\noindent\textbf{Protecting data privacy.}
Recall that SecEL uses the secret hidden in the symmetric bivariate polynomial $\textsf{F}(x,y)$ to mask the sharing parameters by one-time padding. In this way, the confidentiality of the sharing parameters is ensured, and then the privacy of corresponding data is protected. Formally, we define the method in SecEL to mask parameters is semantically secure as shown in \textit{Lemma}~\ref{lemma:1}.

\begin{lemma}\label{lemma:1}
If $\textsf{PRG}$ is a secure pseudo-random function (PRF), SecEL is semantically secure according to the \textit{Definition 1} as shown in the work~\cite{tran2016efficient}
\end{lemma}

\noindent\textbf{Analysis.} Refer it to the work~\cite{tran2016efficient}'s \textit{Definition 1} which defines the security model of one-time padding by games between \textit{challenger} and \textit{adversary}. We obtain the following conclusion: the probability that an adversary correctly guesses a masking parameter $Pr_m$ is less than the probability $Pr_{prg}$ to successfully distinguish $\textsf{PRG}$ and a truly random function plus $\frac{1}{2}$, i.e.,
$Pr_{m} \leq Pr_{prg} + \frac{1}{2}$.

\noindent\textbf{Guaranteeing computation integrity.} Recall that SecEL uses homomorphic authenticator to generate homomorphic message authentication codes for each AV's masking parameters, thereby the computation integrity is guaranteed. Formally, we define the method to generate authentication code is unforgeable as depicted in \textit{Lemma}~\ref{lemma:2}.

\begin{lemma}\label{lemma:2}
If $\textsf{PRG}$ is a secure pseudo-random function (PRF), SecEL is unforgeable according to the work~\cite{tran2016efficient}'s \textit{Definition 2}.
\end{lemma}

\noindent\textbf{Analysis.} Refer it to the work~\cite{tran2016efficient}'s \textit{Definition 2} which defines the security model of generating authenticated code by games between \textit{challenger} and \textit{adversary}. We draw the conclusion: the probability $Pr_{forge}$ that an adversary successfully forges a correct authenticated code is less than the probability ${Pr_{prg}}$ plus the negligible probability for the adversary to make verification queries, i.e., $Pr_{forge} \leq Pr_{prg} + neg(\lambda)$. Herein, $\lambda$ is the security parameter.

\noindent\textbf{Ensuring reliability.} Recall that reliability is referred to allowing for honest AVs correctly decrypting the aggregated parameters only if they have contributions. Refer to the Decryption phase in Section~\ref{sec:design}, we demonstrate how an honest but failed AV's secret shares can be recovered. Particularly, the AV makes his contribution in Aggregation phase but loses his secret shares due to the failed connection in Setup phase. Let $t$ alive AVs help the honest but failed AV to obtain his shares. Then the AV fluently decrypts the aggregated result, thereby avoiding hindering his local training process, which ensures reliability.

\section{Simulation and Evaluation}\label{sec:implementation}
In this section, we simulate and evaluate the presented SecEL. We apply it into a popular image classification task. Extensive experiments are performed to validate the feasibility and effectiveness of SecEL.
\subsection{Simulation Environment}
We simulate SecEL as a module using Python programming language (version 3.6.4) and PyCryptodome library (version 3.6.1), about 200-line codes at total.
The module covers five phases presented in Section \ref{sec:design}.
Note that the length of randomly selected secret is 128~bits and $p$ is a randomly chosen 130~bit prime.
On the other hand, we build the learning model on the MNIST dataset, which is implemented by Python as well, Numpy (version 1.14.0) and Tensorflow (version 1.7.0).
The learning model is collaboratively trained by multiple parties via sharing individual model parameters (also called as gradients).
Herein, the party refers to the role of AV.
Particularly, MNIST dataset is split equally and distributed to individual parties before learning.
Individual parties train a common learning model based on the individual dataset.
At the same time, they share model parameters obtained from their local models.
Shared parameters are masked by calling the \textit{mask} function of the SecEL module.
Those masked parameters then are aggregated by calling SecEL's \textit{aggregate} function.
With \textit{verify} and \textit{decrypt} functions, the aggregated result can be verified and decrypted.
In addition, all of the experiments are conducted on a desktop computer with 2.70\;GHz Intel(R) Xeon(R) CPU and 8\;GB memory.
\subsection{Performance Analysis}
According to the detailed design demonstrated in Section~\ref{sec:design}, we give the performance analysis for SecEL in each round, in terms of computation and communication overhead.
\subsubsection{Computation overhead}
First, for each AV, there has $\mathcal{O}(Nt+l+N+Ntl)$ computation time cost at total, where $l$ is the length of $\mathbf{w}$.
In detail, this total cost can be broken up into the following parts: (i) In Setup phase, each AV needs to share secret to other AV by using Algorithm~\ref{algo:secretsharing}, which leads to time cost $\mathcal{O}(N(t-1))$;
(ii) In Masking phase, each of them masks parameters by the encryption method of one-time padding, which results in time cost $\mathcal{O}(l)$;
(iii) In Verification phase, each active AV collaborates with others to recover the authenticated key and then verifies the returned result, which has time cost $\mathcal{O}(Nl)$;
(iv) In Decryption phase, each active AV needs to rebuild the masking secret and unmask the returned ciphertext with time cost $\mathcal{O}(Ntl)$.
Second, for the edge node, it has time cost $\mathcal{O}(Nl)$, since it needs to aggregate parameters sent by participating AVs.
\begin{table}[htbp]
\scriptsize
 \caption{Performance analysis on computation overhead}\label{tab:computation}
 \begin{threeparttable}
 \begin{tabular}{p{15pt}cp{25pt}cp{20pt}cp{25pt}cp{25pt}lp{40pt}r}
  \toprule
   & \textbf{Setup} & \textbf{Masking} & \textbf{Aggregation} & \textbf{Verification} & \quad\textbf{Decryption}\\
  \midrule
 \textbf{Each AV} & $\mathcal{O}(N(t-1))$ & $\mathcal{O}(l)$ & - - & $\mathcal{O}(Nl)$ & $\mathcal{O}(Ntl)$\\
 \\
 \textbf{Edge node} & - -  & - - &  $\mathcal{O}(Nl)$ &  - - & - -\\
  \bottomrule
 \end{tabular}
  \end{threeparttable}
\end{table}
\subsubsection{Communication overhead}
First, AVs have two kinds of communication cost depending on whether being the leader or not in Verification and Decryption phase.
Recall that we use a randomness algorithm to select a leader among active AVs to help recovering secret, which takes $\mathcal{O}(c^2N)$ communication cost according to the latest work \cite{syta2017scalable}; if $c$ is far smaller that $N$, it would lead to $\mathcal{O}(N)$.
Next, we analyse the communication cost for the AV being leader and the AV not being leader.
On one hand, the AV being leader has $\mathcal{O}(Nt+l+Nl+c^2N+N^2+N^2)$ cost, reduced to $\mathcal{O}(Nl+N^2)$.
On the other hand, the AV not being leader has $\mathcal{O}(Nt+l+Nl+c^2N+N+N)$, reduced to $\mathcal{O}(Nt+Nl)$.
Second, for the edge node, it takes $\mathcal{O}(Nl)$ when receiving the masking parameters from $N$ AVs.
\begin{table}[htbp]
\scriptsize
 \caption{Performance analysis on communication overhead}\label{tab:communication}
 \begin{threeparttable}
 \begin{tabular}{p{20pt}cp{25pt}cp{20pt}cp{25pt}cp{35pt}cp{35pt}c}
  \toprule
   & \textbf{Setup} & \textbf{Masking} & \textbf{Aggregation} & \textbf{Verification} & \quad\quad\quad\textbf{Decryption}\\
  \midrule
 \textbf{Each AV} & $\mathcal{O}(Nt)$ & $\mathcal{O}(l)$ & $\mathcal{O}(Nl)$ & $\mathcal{O}(c^2N+N^2)$ & $\mathcal{O}(N^2)$\\
                  &         &        &         &  or $\mathcal{O}(N)$    & or $\mathcal{O}(N)$\\
 \\
 \textbf{Edge node} & - -  & - - &  $\mathcal{O}(Nl)$ &  - - & - -\\
  \bottomrule
 \end{tabular}
  \end{threeparttable}
\end{table}
\subsection{Evaluation}
We first conduct the experiments to evaluate SecEL's performance from the aspects of time cost and throughput. Time cost is an essentially crucial metric in the setting of AVNET that indicates processing latency. Throughput also is an important factor to demonstrate the consumption of bandwidth resource.
We evaluate the time cost of each phase respectively and throughput in Masking and Aggregation phase.
\begin{figure}[htbp]
\centering\includegraphics[width=5cm,height=4cm]{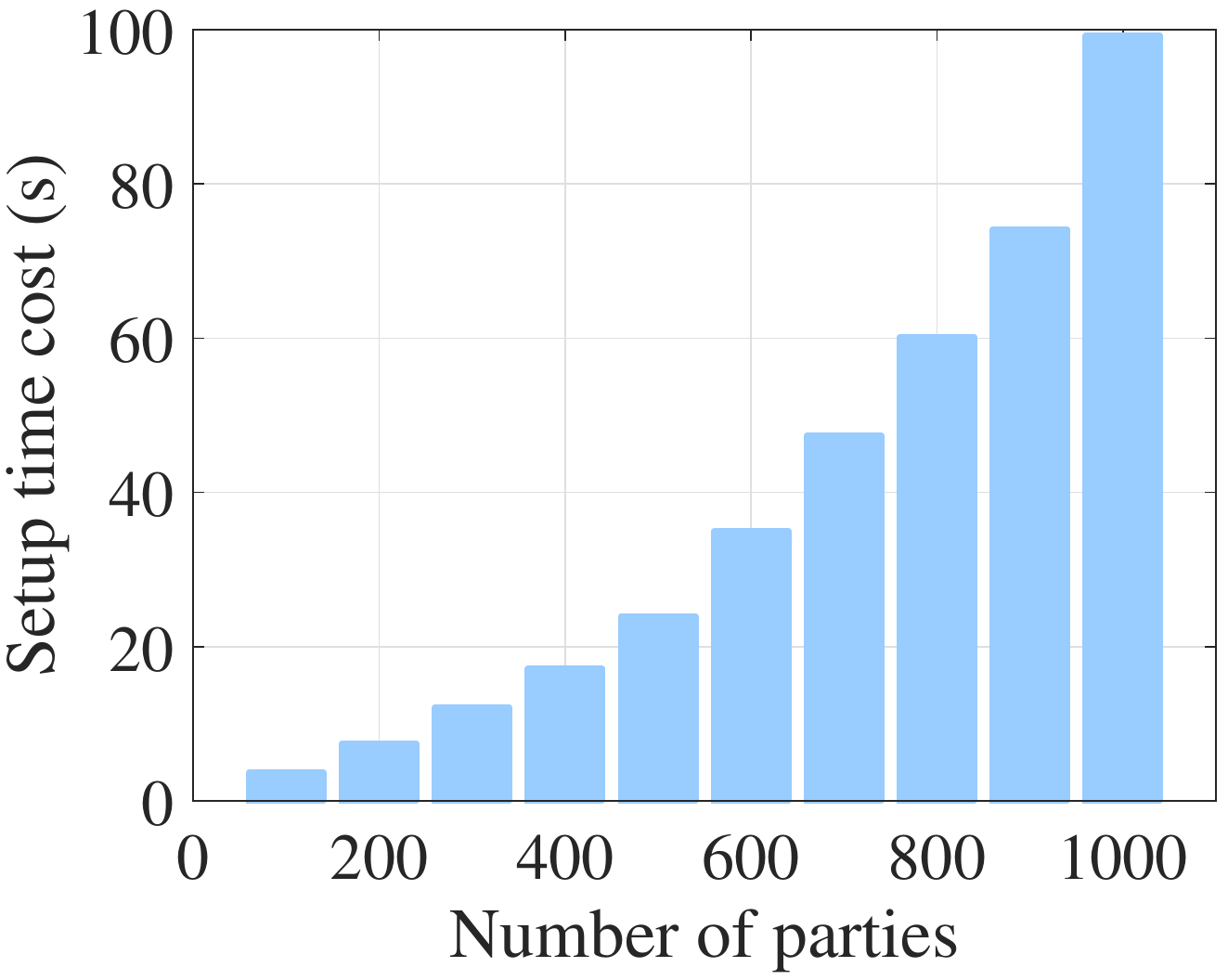}
\caption{Impact of No. of parties on time cost in Setup phase.}
\label{fig:setup}
\end{figure}
\begin{figure}[htbp]
    \centering
    \subfloat[]{        \label{fig:encrypt}
        \begin{minipage}[c]{0.45\linewidth}
            \centering
            \includegraphics[width=1.0\textwidth]{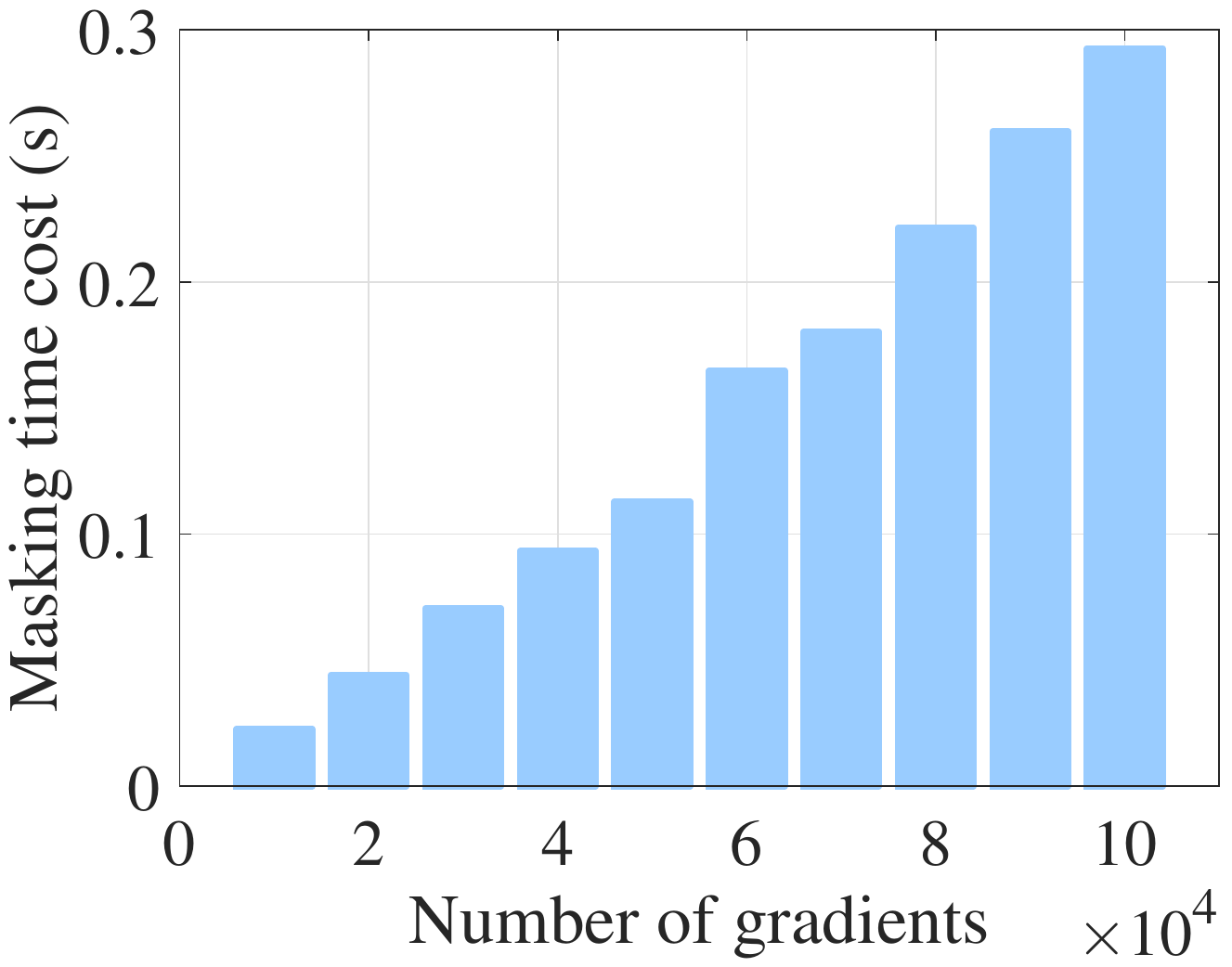}
        \end{minipage}
    }
    \subfloat[]{       \label{fig:encthroughput}
        \begin{minipage}[c]{0.45\linewidth}
            \centering
            \includegraphics[width=1.0\textwidth]{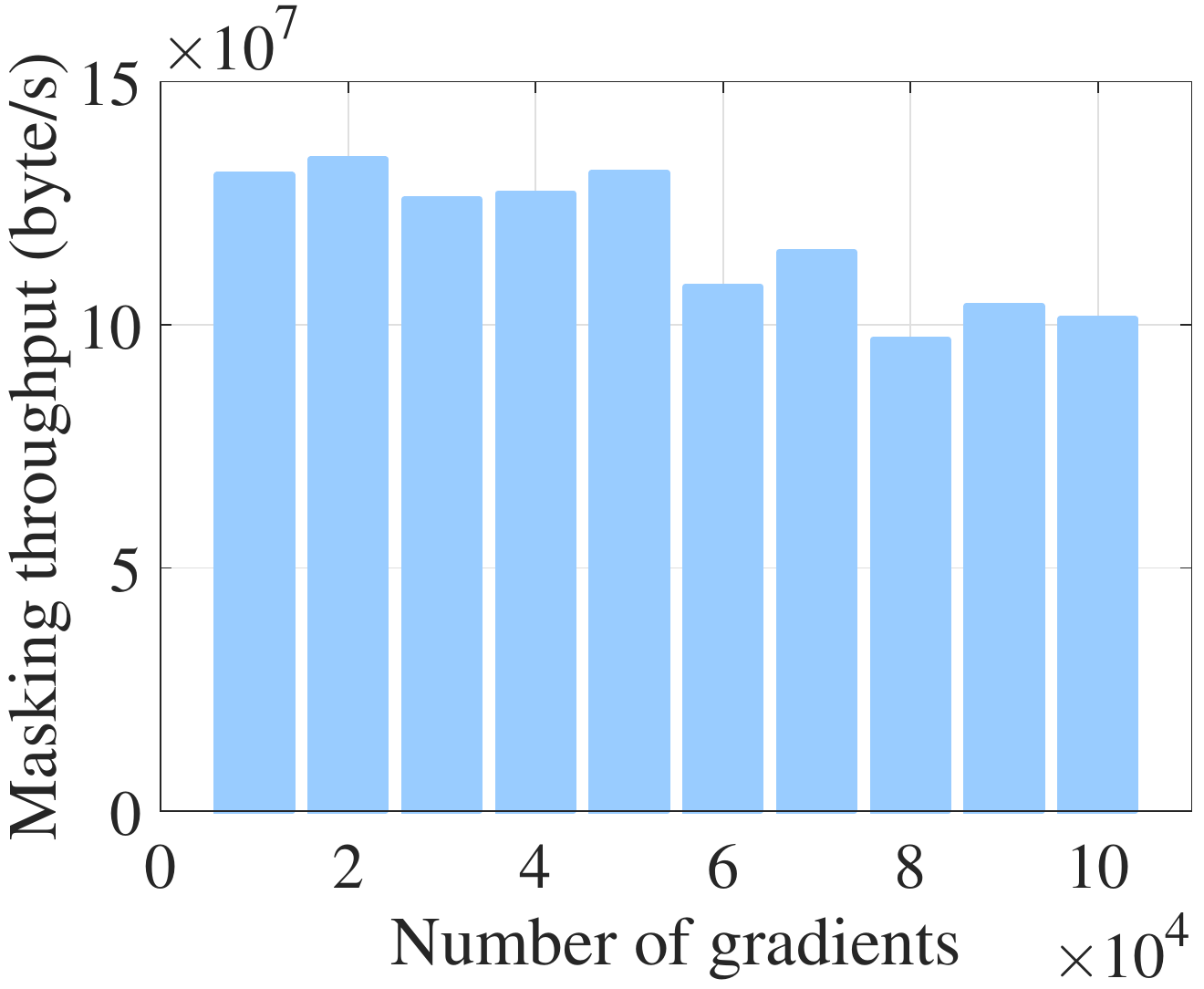}
        \end{minipage}
    }
    \caption{Impact of No. of gradients on time cost and throughput in Masking phase.}
\end{figure}

Fig.~\ref{fig:setup} shows the time consumption in Setup phase with the increasing number of parties.
It is worth noting that the consumed time is independent of the size of learning model, i.e., the number of gradients in the following text. When the number of parties is less than $400$, it will take no more than $20$\;s. Especially, the time cost for 100 parties is actually $3.638106206$\;s.

Fig.~\ref{fig:encrypt} shows the time cost in Masking phase as the number of gradients grows up.  It can be observed that the time cost increases linearly with the number of gradients. However, it is irrelative to the number of parties, since each party masks gradients individually. On the other hand, the trend of throughput appears to be slightly downward as depicted in Fig.~\ref{fig:encthroughput}.

\begin{figure}[htbp]
    \centering
    \subfloat[]{        \label{fig:aggregation}
        \begin{minipage}[c]{0.45\linewidth}
            \centering
            \includegraphics[width=1.0\textwidth]{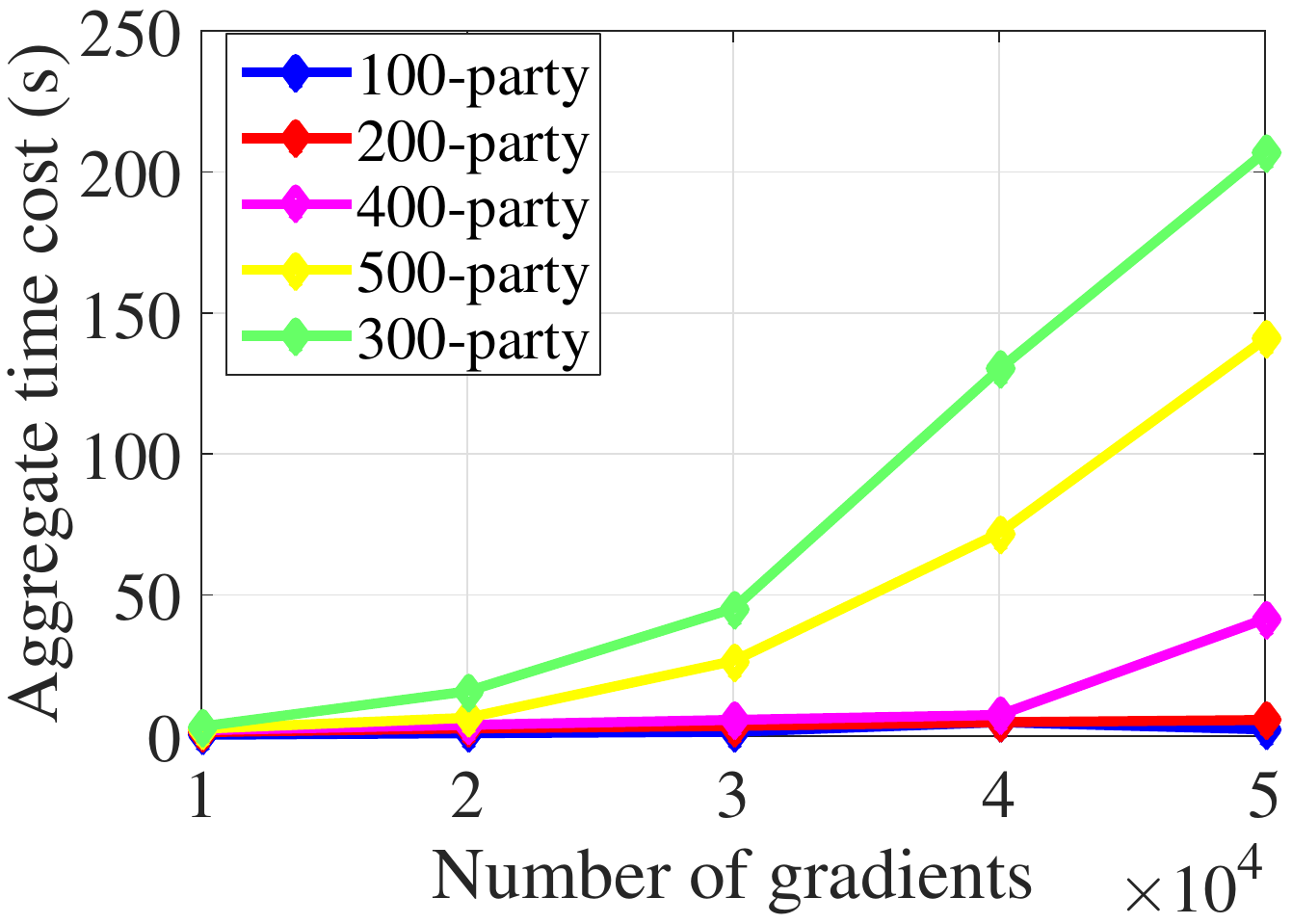}
        \end{minipage}
    }
    \subfloat[]{       \label{fig:aggthroughput}
        \begin{minipage}[c]{0.45\linewidth}
            \centering
            \includegraphics[width=1.0\textwidth]{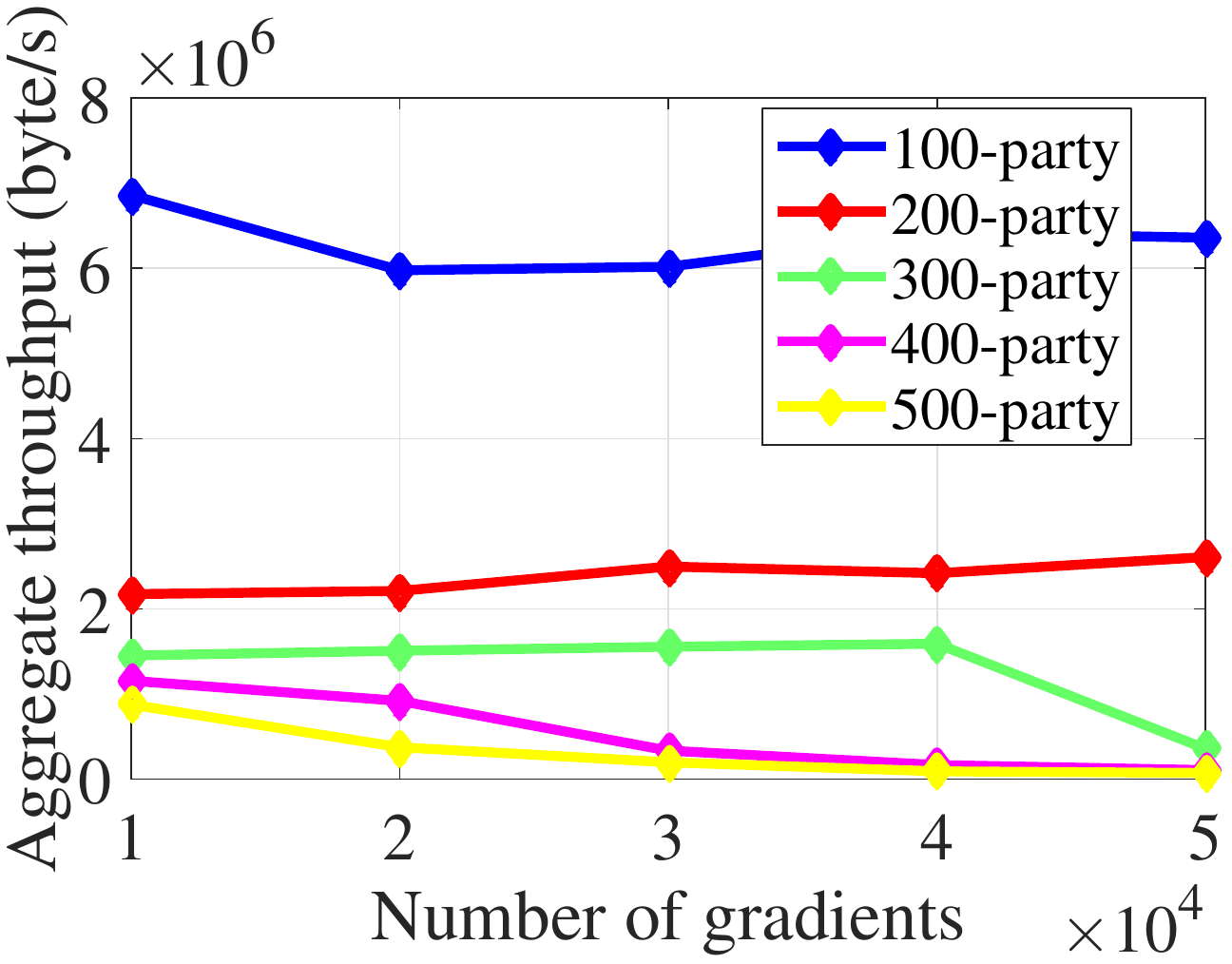}
        \end{minipage}
    }
    \caption{Impact of No. of gradients and No. of parties on time cost and throughput in Aggregation phase.}
\end{figure}

 Fig.~\ref{fig:aggregation} illustrates the time consumption in Aggregation phase, which depends on the quantity of gradients and parties.
It can be observed that as the number of gradients increases, it becomes more time consuming when a group with the same number of parties performs in Aggregation phase. On the other hand, Fig.~\ref{fig:aggthroughput}
 presents that throughput has a decreased tread with more parties participating in.

\begin{figure}[htbp]
    \centering
    \subfloat[]{        \label{fig:verification}
        \begin{minipage}[c]{0.45\linewidth}
            \centering
            \includegraphics[width=1.0\textwidth]{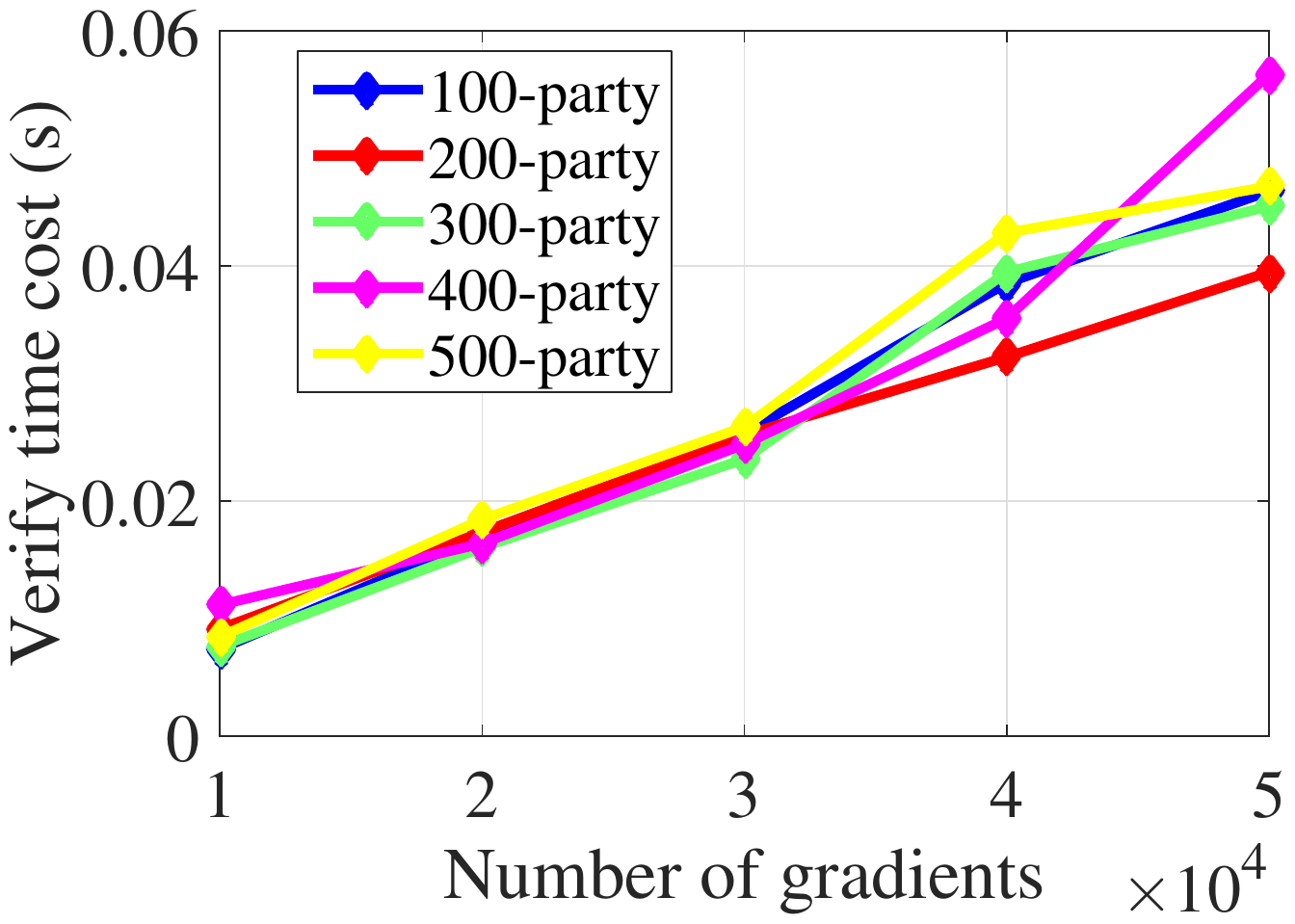}
        \end{minipage}
    }
    \subfloat[]{       \label{fig:decrypt}
        \begin{minipage}[c]{0.45\linewidth}
            \centering
            \includegraphics[width=1.0\textwidth]{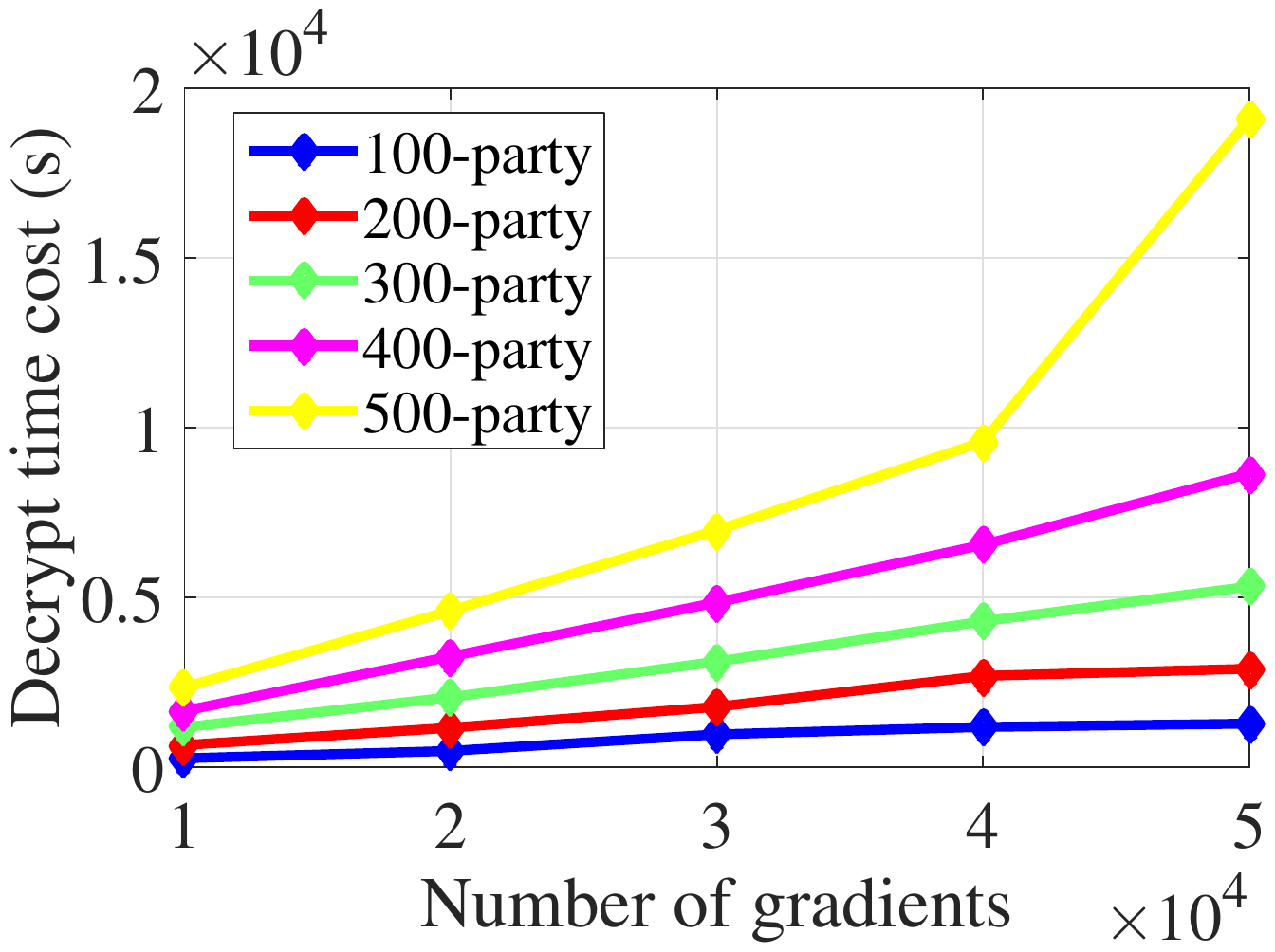}
        \end{minipage}
    }
    \caption{Impact of No. of gradients and No. of parties on time cost in Verification phase and Decryption phase.}
\end{figure}

It can be observed from Fig.~\ref{fig:verification} that the number of parties makes insignificant impact on the time consumed in Verification phase, but the rising number of gradients leads to the linearly increasing time consumption. It is reasonable that the figure of parties is far less than that of gradients and more aggregated gradients take more time to verify.
Finally, it can be observed from Fig.~\ref{fig:decrypt} that the Decryption phase is  more time consuming when compared with other phases due to the overhead of reconstructing masking keys.  As demonstrated, the used time grows up on two hands, i.e., the rising number of gradients and parties.

\begin{figure}[htbp]
\centering\includegraphics[width=5cm,height=4cm]{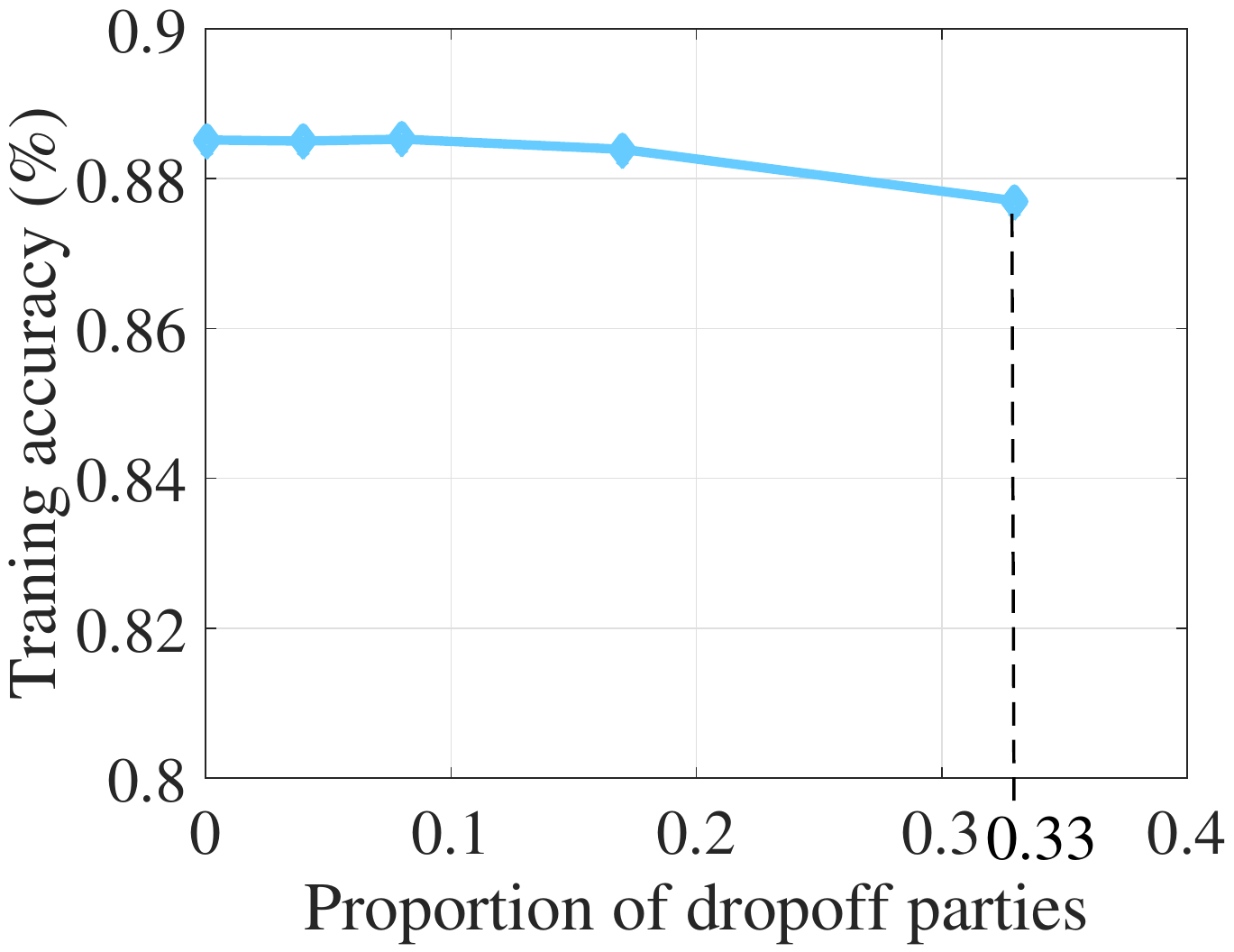}
\caption{Impact of dropoff parties on classification accuracy.}
\label{fig:accuracy}
\end{figure}

After showing each phase's time-cost performance of SecEL, we following demonstrate the impact of different proportions of drop-off parties $f$ on classification accuracy, when the total number of parties is $100$.
In detail, we randomly select $f= \frac{1}{24}, \frac{1}{12}, \frac{1}{6}, \frac{1}{3}$ of $100$ parties failing to share their parameters in each global aggregation step when training. Fig.~\ref{fig:accuracy} gives the experiment results, where the classification accuracy sightly decreases as the figure of drop-off parties grows up.

After showing the time-cost performance of SecEL, we following demonstrate the accuracy results of the classification task in the cases of different number of participating parties with $f=t-1$ parties dropoff. Specifically, we randomly select $f$ parties failing to share their parameters in each global aggregation step, in which $f$ equals to $\frac{1}{3}$ of the total participating parties.
\section{Related Work}\label{sec:related}
\begin{table}[htbp]
 \caption{Comparison of our work and the existing work}\label{tab:comparison}
 \begin{threeparttable}
 \scriptsize
 \begin{tabular}{p{67pt}p{25pt}p{30pt}p{35pt}p{35pt}}
  \toprule
   & \textbf{Ours} & \textbf{\cite{bonawitz2017practical}} & \textbf{\cite{aono2018privacy}} & \textbf{\cite{weng2018deepchain}}\\
  \midrule
 \textbf{Encryption method} & one-time padding & one-time padding & asymmetric encryption & asymmetric encryption \\
 \\
 \textbf{Data privacy} & $\surd$  & $\surd$ &  $\surd$ &  $\surd$\\
 \\
 \textbf{Verifiable computation} & $\surd$ & $\times$ & $\times$ & $\surd$ \\
 \\
 \textbf{Lost secret recovery} & $\surd$ & $\times$ & $\times$ & $\times$\\
  \bottomrule
 \end{tabular}
 \begin{tablenotes}
     \item[1] $\surd$ means having the corresponding security property while $\times$ means not having.
  \end{tablenotes}
  \end{threeparttable}
\end{table}
Recently, there is a wide and hot discussion on edge learning.
Some works study the performance of edge learning, e.g., accuracy and efficiency. For example, Wang \textit{et al.}~\cite{wang2018edge} focus on the convergence rate of the gradient-descent based distributed learning algorithms in resource-limited MEC systems. The authors present an effective control algorithm after analyzing the trade-off relationship among the number of global aggregation, the classification accuracy and resource cost. Specifically, the control algorithm is to make the best use of given amount of resources to learn by balancing well the number of local update and global aggregation.
Also, Zhu \textit{et al.}~\cite{zhu2018towards} research on the communication latency issue of distributed learning in MEC settings and propose the novel concept of learning-driven communication. In a word, the foregoing work can be regarded as the complementary parts to this paper.

On the other hand, to the best of our knowledge, there still have few papers discussing security issues of edge learning. Although there exists a portion of work on privacy-preserving distributed learning, they could not be directly immigrated into the MEC systems, where communication is asynchronous and resources are limited.
We compare these works in Table~\ref{tab:comparison}, which demonstrates SecEL simultaneously achieves three security goals. First of all, Shokri \textit{et al.}~\cite{shokri2015privacy} implement a privacy-preserving distributed deep learning system, where multiple parties share a small fraction of gradients and learn a deep learning model together. Their system uses the difference privacy technique to add noise into shared parameters, so that data privacy is guaranteed to a certain degree. However, their proposed system has been attacked by Hitaj
\textit{et al.}~\cite{hitaj2017deep} by employing the tool of GAN (Generative Adversarial Network).
Bonawitz \textit{et al.}~\cite{bonawitz2017practical} preserve Shokri \textit{et al.}`s system model of training and proposes an efficient method to securely aggregate local gradients of participants for a common deep learning model. However, the presented scheme is deployed in the synchronous communication environment and could not help a fraction of link-failed participants to recover their lost secrets. Most unfortunately, the scheme does not support participants to verify the correctness of computation in a server (the server is the edge node in our paper). As for data privacy, Bonawitz \textit{et al.}'s scheme allows the server to obtain the unmasking aggregated result, while our work does not do that. In SecEL, the masking aggregated result is returned by the edge node and unmasked by participating AVs. Other works on privacy-preserving distributed learning, such as \cite{aono2018privacy, weng2018deepchain}, utilize public key-based cryptographic systems to protect data privacy, which takes more computation than the encryption method of one-time padding.
\section{Conclusion}\label{sec:conclusion}
In this paper, we present a privacy-preserving, verifiable and fault-tolerant scheme, named SecEL for edge learning in AVNET. Specifically, SecEL combines the primitive bivariate polynomial-based secret sharing with homomorphic authenticator. Participating AVs' sharing parameters are protected by the way of one-time padding and labeled with respective MACs, thus data privacy and verifiable computation are ensured. In addition, SecEL allows an honest but failed participating AV's secret can be rebuilt by other active participants, which is adapted to the asynchronous AVNET environment.
Finally, the evaluation of SecEL demonstrates the acceptable performance results.

\section*{Acknowledgement}
Jian Weng was supported by National Natural Science Foundation of China (Grant Nos. 61825203, U1736203 and No. 61732021), Guangdong Provincial Special Funds for Applied Technology Research and Development and Transformation of Key Scientific and Technological Achievements (Grant No. 2016B010124009), and Science and Technology Program of Guangzhou of China (Grant No. 201802010061). Jiasi Weng was supported by National Key R\&D Program of China (Grant No. 2018YFB1402600). Yue Zhang was supported by National Natural Science Foundation of China (Grant No. 61872153). Ming Li was supported by National Key Research and Development Plan of China (Grant No. 2017YFB0802203), and Graduate School of Jinan University.




\bibliographystyle{IEEEtran}
\bibliography{SecEL}
\begin{IEEEbiography}[{\includegraphics[width=1in,height=1.25in,clip,keepaspectratio]{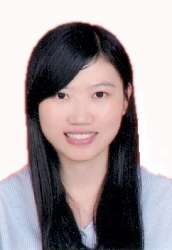}}]
{Jiasi Weng} received the B.S. degree in software engineering from South China Agriculture University in June 2016. Currently, she is a Ph.D. student in the Colledge of Information Science and Technology \& College of Cyber Security at Jinan University. Her research interests include applied cryptography, blockchain and privacy.
\end{IEEEbiography}
\begin{IEEEbiography}[{\includegraphics[width=1in,height=1.25in,clip,keepaspectratio]{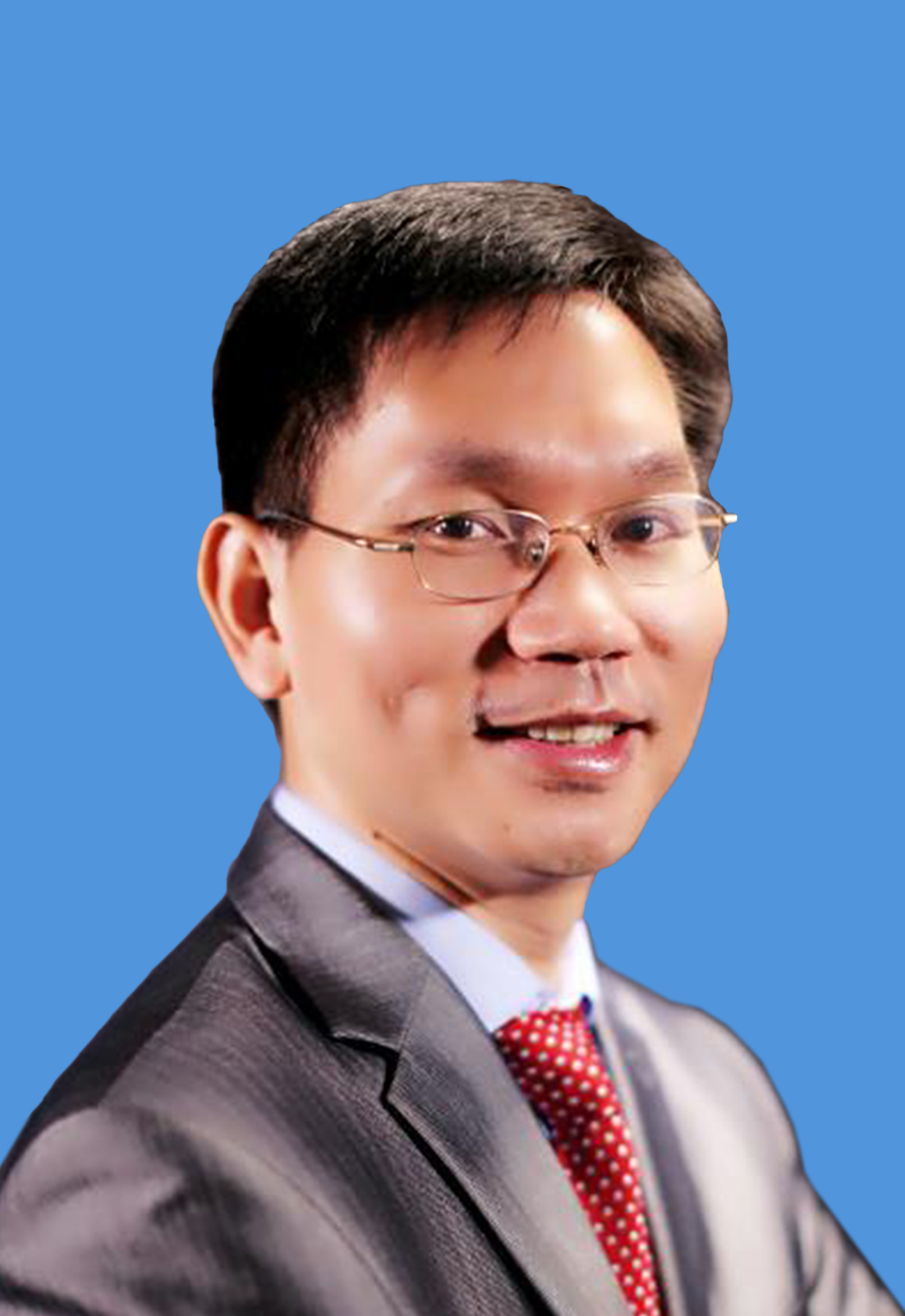}}]
{Jian Weng} is a professor and the Executive Dean with College of Information Science and Technology in Jinan University. He received B.S. degree and M.S. degree from South China University of Technology in 2001 and 2004 respectively, and Ph.D. degree at Shanghai Jiao Tong University in 2008. His research areas include public key cryptography, cloud security, blockchain, etc. He has published 80 papers in international conferences and journals such as CRYPTO, EUROCRYPT, ASIACRYPT, TCC, PKC, CT-RSA, IEEE TPAMI, IEEE TDSC, etc. He also serves as associate editor of IEEE Transactions on Vehicular Technology. He received the Young Scientists Fund of the National Natural Science Foundation of China in 2018, and the Cryptography Innovation Award from Chinese Association for Cryptologic Research (CACR) in 2015. He served as General Co-Chair for SecureComm 2016, TPC Co-Chairs for RFIDsec'13 Asia and ISPEC 2011, and program committee members for more than 40 international cryptography and information security conferences. He also serves as associate editor of IEEE Transactions on Vehicular Technology.
\end{IEEEbiography}
\begin{IEEEbiography}[{\includegraphics[width=1in,height=1.25in,clip,keepaspectratio]{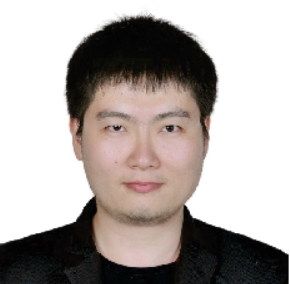}}]
{Yue Zhang} is a Ph.D. student in the Colledge of Information Science and Technology \& College of Cyber Security at Jinan University. His research focuses on system security, especially IoT security. He has published papers in international conference and journals, such as IEEE TDSC, IEEE TPDS, IEEE TVT, RAID, etc.
\end{IEEEbiography}
\begin{IEEEbiography}[{\includegraphics[width=1in,height=1.25in,clip,keepaspectratio]{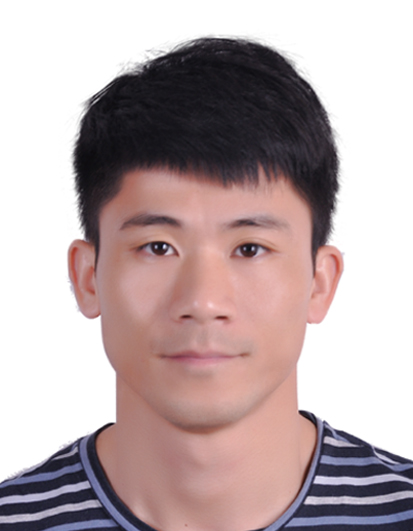}}]
{Ming Li} received his B.S. in electronic information engineering from University of South China in 2009, and M.S. in information processing from Northwestern Polytechnical University in 2012. From 2016, he becomes a Ph.D. student in the Colledge of Information Science and Technology \& College of Cyber Security at Jinan University. His research interests include crowdsourcing, blockchain and its privacy and security.
\end{IEEEbiography}
\end{document}